\documentclass[aps,prl,twocolumn,showpacs,superscriptaddress]{revtex4}
\usepackage{graphicx}
\usepackage{bm}

\begin{document}

%\preprint{}
\title{Uniaxial $c$-axis pressure effects on underdoped BaFe$_2$(As$_{0.72}$P$_{0.28}$)$_2$ superconductor}

\author{Ding Hu}
\email{dinghuphys@gmail.com}
\affiliation{Center for Advanced Quantum Studies and Department of Physics, Beijing Normal University, Beijing 100875, China}
\affiliation{Department of Physics and Astronomy, Rice University, Houston, Texas 77005-1827, USA}
\author{David W. Tam}
\affiliation{Department of Physics and Astronomy, Rice University, Houston, Texas 77005-1827, USA}
% experiment at HB-1A
\author{Wenliang Zhang}
\affiliation{Beijing National Laboratory for Condensed Matter Physics, Institute of Physics, Chinese Academy of Sciences, Beijing 100190, China}
%\affiliation{School of Physical Sciences, University of Chinese Academy of Sciences, Beijing 100190, China}
% sample growth at IOP
\author{Yuan Wei}
\affiliation{Beijing National Laboratory for Condensed Matter Physics, Institute of Physics, Chinese Academy of Sciences, Beijing 100190, China}
%\affiliation{School of Physical Sciences, University of Chinese Academy of Sciences, Beijing 100190, China}
\author{Robert Georgii}
\affiliation{Heinz Maier-Leibnitz Zentrum, Technische Universit$\ddot{a}$t M$\ddot{u}$nchen, D-85748 Garching, Germany}
% local contact of FRM-II
\author{Bj$\rm \ddot{o}$rn Pedersen}
\affiliation{Heinz Maier-Leibnitz Zentrum, Technische Universit$\ddot{a}$t M$\ddot{u}$nchen, D-85748 Garching, Germany}
% local contact of FRM-II
\author{Alfonso Chacon Roldan}
\affiliation{Physik-Department, Technische Universit$\ddot{a}$t M$\ddot{u}$nchen, D-85748 Garching, Germany}
\author{Pengcheng Dai}
\email{pdai@rice.edu}
\affiliation{Department of Physics and Astronomy, Rice University, Houston, Texas 77005-1827, USA}
\affiliation{Center for Advanced Quantum Studies and Department of Physics, Beijing Normal University, Beijing 100875, China}

\begin{abstract}
The optimal superconductivity ($T_c \approx 30 $ K) in BaFe$_2$(As$_{1-x}$P$_x$)$_2$ can be reached
when the coupled antiferromagnetic (AF) order ($T_N$) and orthorhombic lattice distortion ($T_s$) are
suppressed to zero temperature with increasing of P concentration or hydrostatic pressure.
Here we use transport and neutron scattering to study
the $c$-axis pressure effects on electronic phases in underdoped BaFe$_2$(As$_{0.72}$P$_{0.28}$)$_2$, which has $T_N = T_s\approx 40$ K and $T_c \approx $ 28 K at zero pressure.
With increasing $c$-axis pressure, $T_N$ and $T_s$ are slightly enhanced around $P_c \sim 20 $ MPa. Upon further increasing pressure, AF order is gradually suppressed to zero, while $T_c$ is enhanced to 30 K.
%optimal superconductivity with $T_c \sim $ 30 K is reached.
Our results reveal the importance of magnetoelastic couplings in BaFe$_2$(As$_{1-x}$P$_x$)$_2$, suggesting that the $c$-axis pressure can be used as a tuning parameter to manipulate the electronic phases in iron pnictides.
%the importance of quantum fluctuation related to the suppression of coupled AF order and nemtaic phase to zero temperature.
\end{abstract}



\maketitle

The parent compounds of iron-based superconductors are long-range ordered antiferromagnets below
a N$\rm \acute{e}$el temperature $T_N$ and
also display tetragonal to orthorhombic lattice distortion below $T_s$ ($T_s\geq T_N$)
\cite{hosono2008,stewart,dai}. High-temperature superconductivity in these materials can be induced by charge carrier doping, chemical pressure, and hydrostatic pressures
that act to suppress $T_N$ and $T_s$ in a manner akin to other unconventional superconductors such as cuprates and heavy-fermions \cite{scalapinormp}. To understand the microscopic origin of superconductivity, it is therefore important to sort out the interplay amongst magnetism, lattice distortion, and superconductivity. Compared with charge carrier doping and chemical pressure via element substitution, which can also cause lattice disorder, hydrostatic and uniaxial pressure can tune
the electronic, magnetic, and superconducting properties of the system without inducing additional lattice disorder
 \cite{Sun2012,Canfield2008,hosonopressure2008,felser2009,tomi12,takehiro2010,Duncan2010,Dhital12,Lu14,Tam,HRMan18}.

BaFe$_2$As$_2$, one of the parent compounds of iron-based superconductors, undergoes a tetragonal to orthorhombic structural transition at $T_s$ and orders in a colinear antiferromagnetic (AF) structure below $T_N$ $(T_N \approx T_s \approx 140 $ K$ )$ \cite{huang2008,Kim2011}. Upon electron or hole doping to form BaFe$_{2-x}T_{x}$As$_2$ ($T$ = Co, Ni) \cite{Luo2012,Lu2013,mcqueeney2011} or Ba$_{1-x}A_x$Fe$_2$As$_2$ ($A$ = K, Na) \cite{Avci2014,bohmer2015}, static AF order is suppressed and exotic magnetic phases
such as incommensurate and $C_4$ magnetic order appear before doping induced optimal superconductivity.
In the case of isoelectronic doped BaFe$_2$(As$_{1-x}$P$_{x}$)$_2$, the structural and AF phase transitions are always coupled and increasing P-doping suppresses $T_s/T_N$ near $x = 0.30$ where the optimal superconductivity is achieved at $T_c \approx 30 $ K \cite{DHu2015,Allred2014}.
The substitution on the arsenic site by the smaller phosphorous atom is regarded as introducing chemical pressure in the system. Magnetic susceptibility measurements under hydrostatic pressure of underdoped BaFe$_2$(As$_{1-x}$P$_{x}$)$_2$ point to a similar superconducting phase diagram with maximum $T_c \approx 30 $ K \cite{Duncan2010,Lina2010}. Given the similar electronic phase diagrams of P-doped and hydrostatic pressured BaFe$_2$(As$_{1-x}$P$_{x}$)$_2$, it would be interesting to test
the effect of uniaxial pressure of the electronic phase diagram of the system
\cite{Lina2010,Shishido2010}.

Previous study on BaFe$_2$(As$_{1-x}$P$_{x}$)$_2$ with in-plane strain added along the orthorhombic axis reveals increased $T_N$ and decreased $T_c$ \cite{fisher2012,bohmer12}, suggesting that the effect
of the in-plane strain is similar to decreasing $x$ by shifting the phase
diagram \cite{fisher2012,bohmer12}. These results are analogous to the effect of an in-plane strain
on electron-doped BaFe$_{2-x}T_{x}$As$_2$ ($T$ = Co, Ni) \cite{Tam,Lu2016}. Since in-plane strain already breaks the $C_4$ symmetry of the tetragonal phase and induces orthorhombic lattice distortion, strain-induced AF order reveals the subtle balance between magnetism and superconductivity.
On the other hand, pressure dependence of the thermodynamic measurements reveal that
a $c$-axis aligned uniaxial strain on BaFe$_2$(As$_{1-x}$P$_{x}$)$_2$
increases $T_c$ and may correspond to an increased P-doped level \cite{bohmer12}.
Therefore, it is surprising that our recent neutron diffraction
and transport measurements on optimally doped BaFe$_2$(As$_{0.70}$P$_{0.30}$)$_2$ found that
a $c$-axis aligned uniaxial pressure can spontaneously induce static stripe AF order with slightly suppressed $T_c$ \cite{Ding2018}.

As the AF order and nematic phase in BaFe$_2$(As$_{1-x}$P$_{x}$)$_2$ disappears in a weakly first-order fashion near optimal superconductivity [Fig. 1(a)], the $T_N$/$T_s$ is sensitive to change of phosphorous concentration near $x = 0.30$ \cite{DHu2015}.
Thus, we choose to carry out transport and neutron diffraction studies on the underdoped compound BaFe$_2$(As$_{0.72}$P$_{0.28}$)$_2$ with $T_N = T_s $ $\approx$ 40 K and $T_c \approx $ 28 K, to further investigate the effect of a $c$-axis pressure
on the electronic properties of BaFe$_2$(As$_{1-x}$P$_{x}$)$_2$ and determine the origin
of the observed quantum critical fluctuation near optimal superconductivity \cite{Matsuda2014,DHu2015,Kuo2016,James2014}.
%To further investigate the effect of a $c$-axis pressure
%on the electronic properties of BaFe$_2$(As$_{1-x}$P$_{x}$)$_2$ and determine the origin
%of the observed quantum critical point near optimal superconductivity [figure. 1(a)] \cite{Matsuda2014,DHu2015,Kuo2016,James2014}, we use transport and neutron diffraction to study
%underdoped superconducting BaFe$_2$(As$_{0.72}$P$_{0.28}$)$_2$ with $T_N \approx T_s $ $\approx$ 40 K and $T_c \approx $ 28 K.
We find that a $c$-axis aligned uniaxial pressure can significantly affect the AF ordering temperature $T_N$, while only slightly modify superconducting transition temperature $T_c$.
As a function of increasing uniaxial pressure $P_c$, $T_N$ and $T_s$ are slightly enhanced at $P_c \sim 20 $ MPa firstly. Then, they are
gradually suppressed to zero  at $P_c \sim 280 $ MPa with $T_c \approx $ 30 K. These results suggest that a $c$-axis aligned pressure can be used as a tuning parameter to  manipulate the complex electronic phases in iron pnictides.

\begin{figure}
\includegraphics[scale=.425]{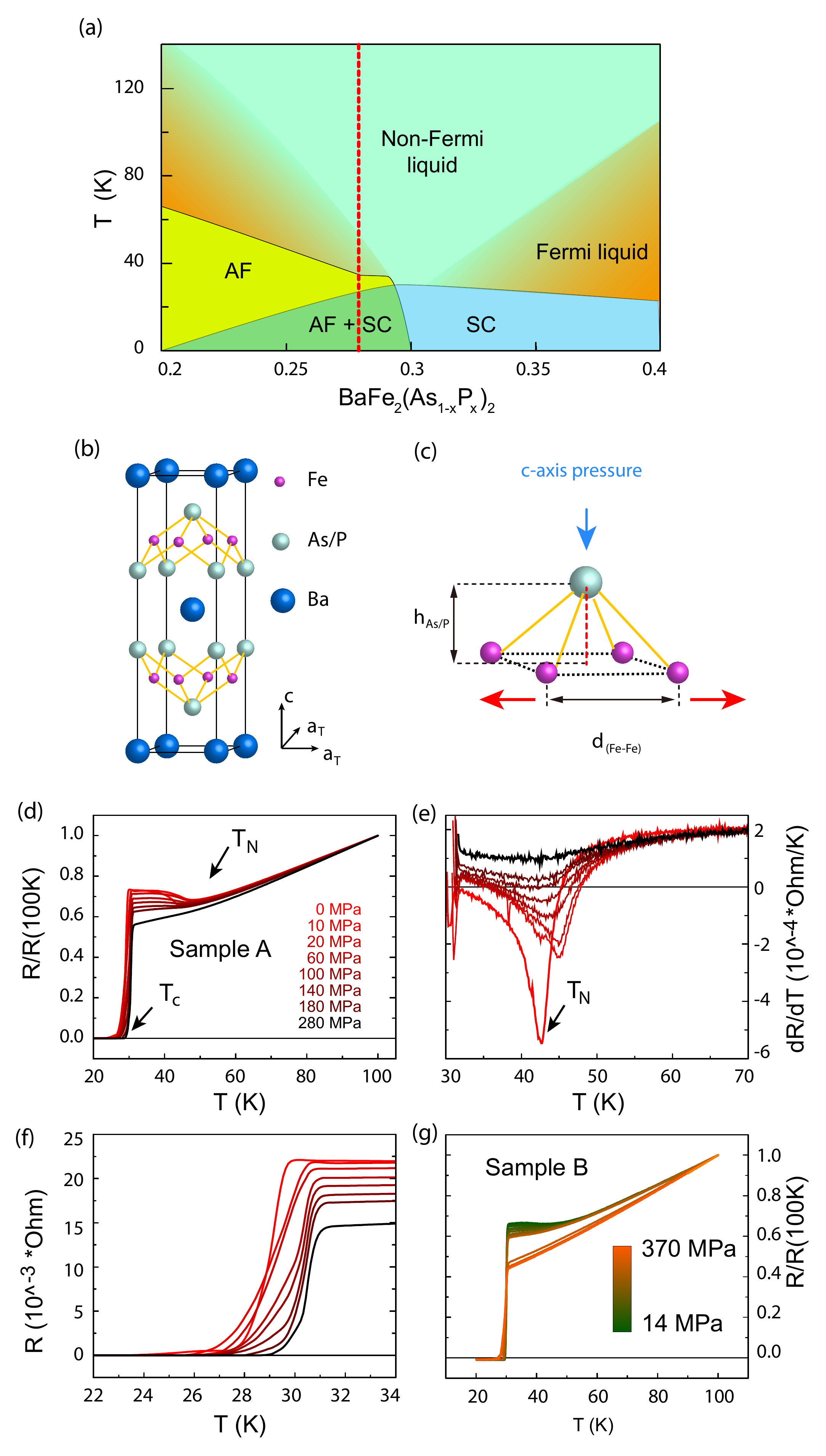}
\caption{(a) The schematic electronic phase diagram of BaFe$_2$(As$_{1-x}$P$_{x}$)$_2$ with $0.2 < x < 0.4$ \cite{Matsuda2014}, where antiferromagnetic order (AF), superconductivity (SC), Fermi liquid, Non-Fermi liquid are marked. The red dashed line marks the position of the $x = 0.28$ compound measured in this work. (b) The crystal structure of BaFe$_2$(As$_{1-x}$P$_{x}$)$_2$. The purple, silvery and blue balls indicate the Fe, As/P, and Ba positions, respectively. (c) The schematic diagram of the FeAs tetrahedron. The red arrow indicates the increasing Fe-Fe distance due to a $c$-axis pressure. (d) Temperature dependence of the in-plane resistance for pressures up-to 280 MPa along the c-axis below 100 K of sample A. Data were normalized to 100 K resistance. (e) The temperature derivative of the in-plane resistance
at different pressure of sample A, which reveals the $T_N$ and $T_s$ more clearly. (f) Temperature dependence of the resistance around 30 K at different pressures of sample A reveal the superconducting transition temperature evolution. (g) Temperature dependence of resistance under c-axis pressures of sample B, which is another single crystal from the same batch of sample A.
}
\label{fig1}
\end{figure}

%While our data measured in ambient condition are consistent with earlier results \cite{DHu2015}, we find that the nearly simultaneous structural and magnetic transitions are enhanced by c-axis pressure firstly (P$_c$ $< 33 $MPa), then suppressed continuously until the abrupt disappearance with reaching the optimal superconductivity (P$_c$ $= 280 $MPa). Surprisingly, when $c$-axis pressure is added further, the structural and magnetic transitions occur again, along with the compression of superconductivity. It resembles the situation found in the optimal single crystal ($x = 0.30$) without a magnetic transition in ambient condition, where the antiferromagnetic order is induced by c-axis pressure. These results point to a universal phase diagram for different phosphorous contents under $c$-axis pressure, suggesting $c$-axis pressure as a new tuning parameter to manipulate the complex electronic phases in iron pnictides.

We chose to study the effects of a $c$-axis aligned uniaxial pressure on slightly underdoped BaFe$_2$(As$_{0.72}$P$_{0.28}$)$_2$ superconductor with
$T_N = T_s \approx 40$ K and $T_c \approx 28$ K, because of its close proximity to optimal superconductivity
but with
distinctively different electronic phases to the $ x$ = 0.30 compound in zero pressure \cite{DHu2015}. The crystal structure of BaFe$_2$(As$_{1-x}$P$_{x}$)$_2$ and its response to a $c$-axis aligned pressure is shown in Figs. 1(b, c)
\cite{Ding2018}.
A custom designed pneumatic uniaxial pressure apparatus was used in the transport measurements, which can control the applied pressure precisely regardless of thermal contraction of the sample and apparatus \cite{David17,Ding2018}. The $c$-axis pressures have been applied successively at 300 K on the same BaFe$_2$(As$_{0.72}$P$_{0.28}$)$_2$ crystal in each measurement.

Figure 1(d) shows temperature dependence of the resistivity at different $c$-axis
pressures up to 280 MPa on sample A.
At zero pressure, we find a kink around 40 K due to the AF order and orthorhombic structure transition \cite{DHu2015}. With increasing $P_c$, the superconducting transition temperature
gradually increases and
reaches maximum ($T_c\approx 30$ K) with the disappearance of the kink above $T_c$ at $P_c = 280$ MPa.
%Upon further increasing pressure $P_c$, $T_c$ starts to decrease and the kink above $T_c$
%reappears, suggesting that a $c$-axis aligned pressure starts to induce AF order.% and lattice distortion.

Assuming that the kink in temperature dependence of the resistivity indeed arises from AF order,
we can determine $c$-axis pressure evolution of the ordering transition $T_N$ by plotting the resistivity derivative $d R/d T$ versus $P_c$ in Fig. 1(e).
Inspection of the figure reveals only one clear dip at each pressure, similar with the results measured in ambient conditions and in-plane strain on BaFe$_2$(As$_{1-x}$P$_{x}$)$_2$ compounds \cite{fisher2012,bohmer12,DHu2015}, suggesting coupled AF order
and orthorhombic lattice distortion at all studied pressures. This is different
from the electron-doped pnictides where two anomalies in $d R/d T$ corresponding to the distinct $T_s$ and $T_N$,  respectively \cite{Rui2014}.

As a function of increasing $P_c$, $T_N$ shown as a dip
in $d R/d T$ increases slightly below 20 MPa, then it
gradually decreases until vanishing at $P_c = 280$ MPa.
%For $P_c\geq 380$ MPa, the dip in $d \rho/d T$ reappears, suggesting the presence of a pressure-induced AF order. This is reminiscent of the $c$-axis pressure
%induced stripe AF order in the optimally doped BaFe$_2$(As$_{0.70}$P$_{0.30}$)$_2$ \cite{Ding2018}.
Figure 1(f) shows the temperature dependence of the resistivity near $T_c$ as a function of $P_c$.
With increasing $P_c$, we see a slight increase in $T_c$ until the maximum $T_c\approx 30$ K is achieved
around 280 MPa. We note that the normal state resistance behavior above $T_N$ at low pressures is different from the curve of 280 MPa. To confirm these results, we carried out the resistance measurements on another single crystal from the same batch marked as sample B. Similar to sample A, we find that uniaxial pressure indeed suppresses the AF order continuously, until it vanishes around 300 MPa [Fig 1(g)].
 %which is consistent with the results from another single crystal with similar concentration (Sfig. 1).   %Upon further increasing $P_c$, $T_c$ decreases smoothly and reaches to $T_c\approx 28$ K at
%$P_c\geq 680$ MPa. The broadening superconducting transition found at $P_c\approx 380$ MPa may be due to the inhomogeneity of $P_c$ and the high sensitivity of $T_c$ to pressure in this pressure regime.
The reduction in $T_N$ and the increased $T_c$ for $20$ MPa $<P_c< 280$ MPa is also seen in
previous work where the value of $P_c$ is unknown \cite{bohmer12}. %In our case, the uniaxial pressure is well calibrated and precisely controlled by
%a pneumatic pressure with a feedback-controlled loop \cite{Ding2018}.

\begin{figure}
\includegraphics[scale=0.45]{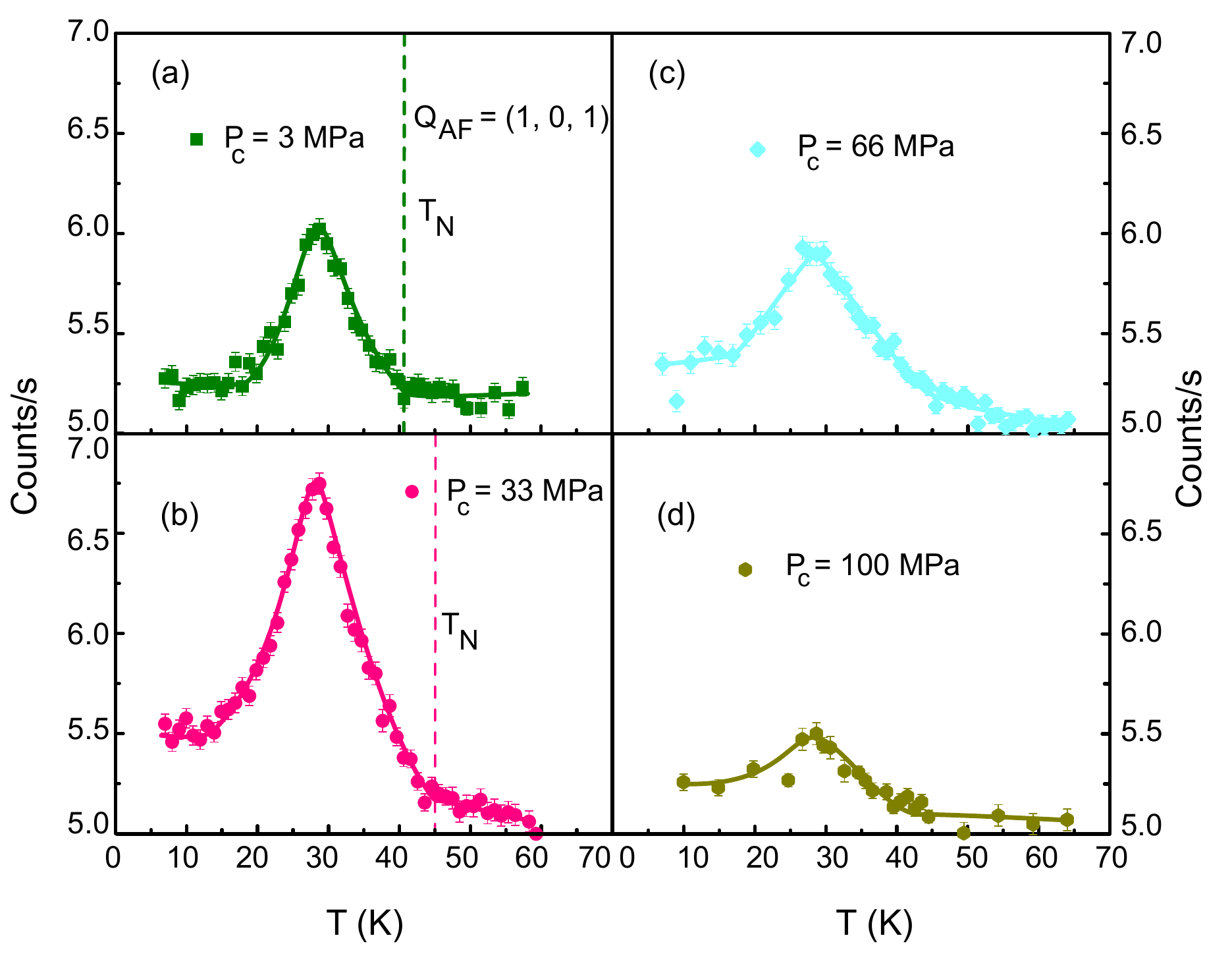}
\caption{Temperature dependence of the magnetic scattering intensity measured on MIRA-II for BaFe$_2$(As$_{0.72}$P$_{0.28}$)$_2$ with a $c$-axis pressure of (a) P$_c = 3$ MPa, (b) 33 MPa, (c) 66 MPa and (d) 100 MPa at $\textbf{Q}_{AF} = (1, 0, 1)$. The solid lines are guides to the eye. Four figures are plotted in the same vertical scale so pressure-induced magnetic scattering change can be directly compared. Dashed lines in (a) and (b) mark the AF transition temperatures $T_N$. %Antiferromagnetic transition become smooth in (c) and (d) and hard to determine T$_N$ from these neutron data.
}
\label{fig2}
\end{figure}

\begin{figure}
\includegraphics[scale=0.75]{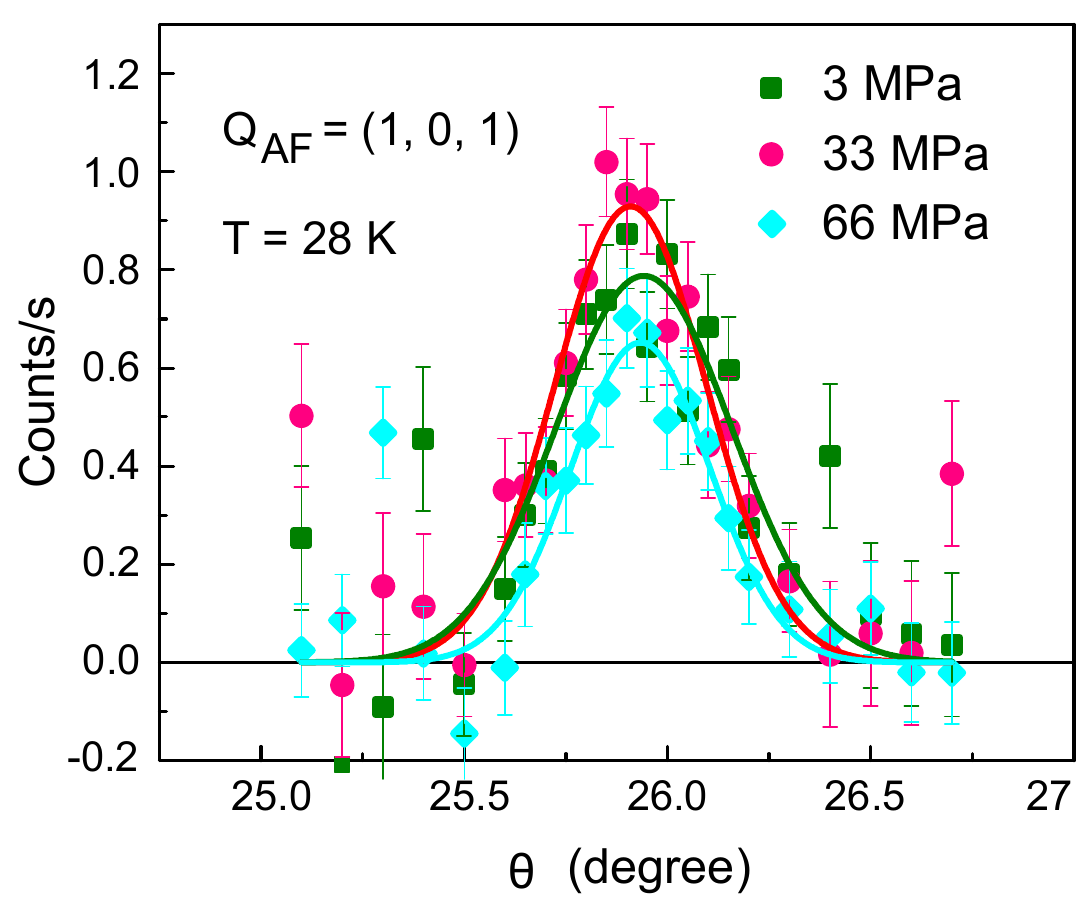}
\caption{Rocking scans across the $\textbf{Q}_{AF} = (1, 0, 1)$ position
measured under different pressures. The solid lines are Gaussian fits to the data. The signals are measured at 28 K with background subtracted. %Horizontal bar indicates instrumental resolution.
The background scattering is measured above $T_N$ but at different temperatures, it may result in intensity comparison of the pressure dependence of the scattering to be different from those of Fig. 2.
}
\label{fig3}
\end{figure}

\begin{figure}
\includegraphics[scale=0.65]{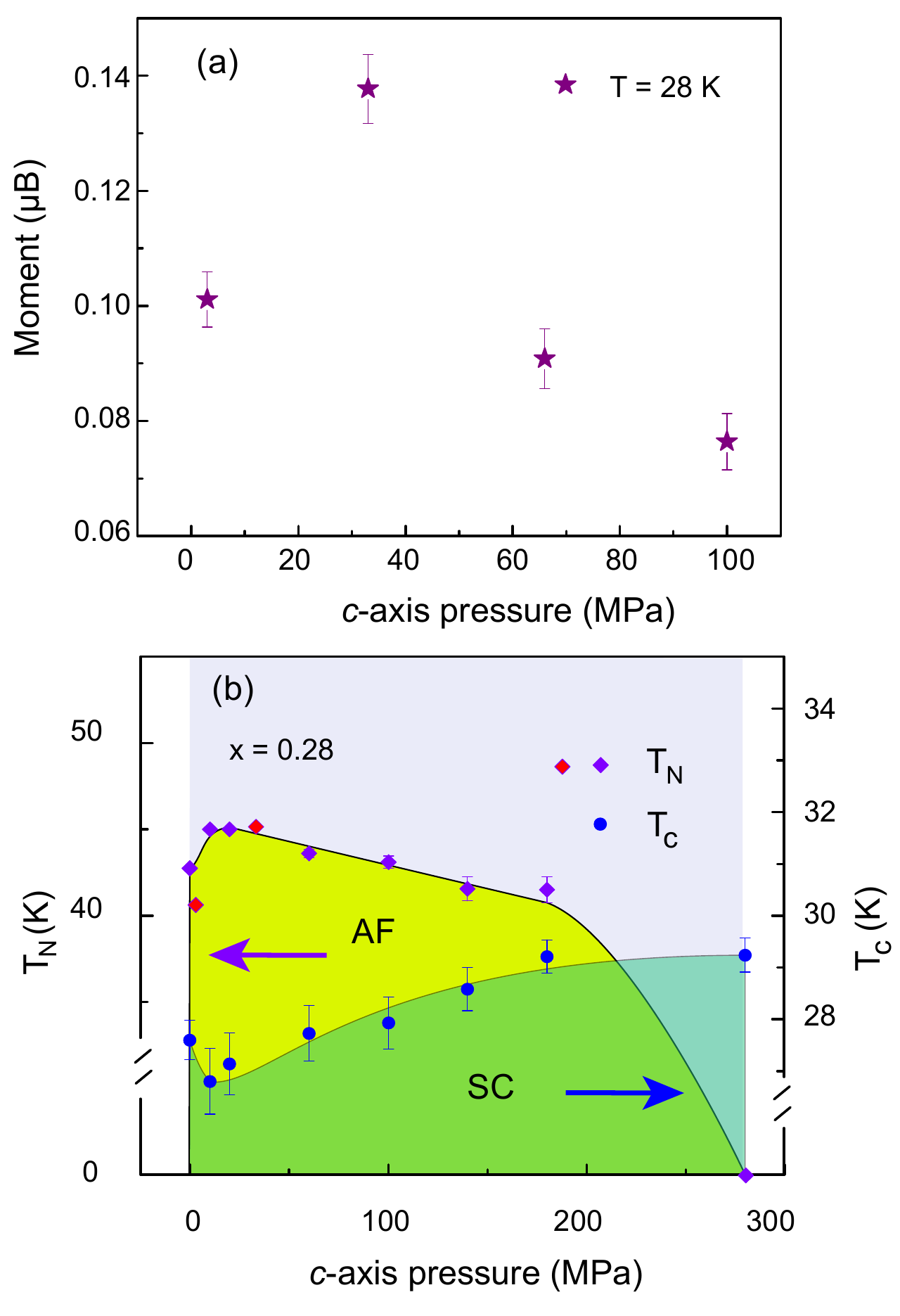}
\caption{(a) The $c$-axis pressure dependence of the ordered magnetic moment at $T = 28$ K estimated assuming a stripe AF structure. The ordered moments are calculated from comparing the magnetic signal intensity from the temperature scans with a weak nuclear structure peak intensity. (b) Electronic phase diagram of BaFe$_2$(As$_{0.72}$P$_{0.28}$)$_2$
as a function of $c$-axis applied pressure extracted from the transport data of Fig. 1.
The diamonds and circles are $T_N$s and $T_c$s corresponding to the
left and right scales, respectively. Red diamonds are $T_N$s obtained from the neutron data.
}
\label{fig4}
\end{figure}

%In previous reports, $c/a$ ratio, the pnictogen height, or the As-Fe-As bond angles are suggested separately to be in connection with the evolution of phase diagram especially the critical superconduting transition temperatures ($T_c$) \cite{mizu10,hosono15,lee08,kuro09,kote08,zhao08,kimber09}.

To confirm the pressure-induced changes in AF order,
we carried out neutron diffraction experiments in the same compound using the MIRA triple-axis spectrometer at Maier-Leibnitz, Garching, Germany \cite{Robert2015,Robert2018}.
The BaFe$_2$(As$_{0.72}$P$_{0.28}$)$_2$ crystal was clamped between two Al plate and loaded in an
in-situ uniaxial pressure cell \cite{Chacon2015}. Due to the large single crystal used to ensure the signal strength in neutron scattering experiments, the in-situ uniaxial pressure cell cannot reach the high pressure limit of the
transport measurements as shown in Fig. 1(d).

Figure 2 summarizes temperature dependence of the scattering at the AF ordering wave vector
$\textbf{Q}_{AF}= (1, 0, 1)$  for pressures up to 100 MPa. At $P_c\approx 3$ MPa, the BaFe$_2$(As$_{0.72}$P$_{0.28}$)$_2$ orders below 40 K [Fig. 2(a)]. The reduction in magnetic scattering below $\approx$ 28 K is due to the appearance of superconductivity \cite{Ding2018}.
Upon application of a $P_c\approx $ 33 MPa pressure, the $T_N$ increases from $\approx$ 40 K to $\approx$ 44 K.
Although it is difficult to determine the $T_N$ precisely from the temperature dependence of the magnetic scattering at 66 MPa and 100 MPa, we can clear see the reduction in magnetic scattering with increasing
pressure. Benefiting from the in-situ pressure cell used in the measurements, we can compare the scattering intensity in Figs. 2(a-d) directly and estimate the magnetic ordered moment at 28 K assuming a stripe AF structure.  The magnetic ordered moment reaches a maximum at 33 MPa and then decreases with further increasing pressure [Figs. 3 and 4(a)], consistent with the evolution of $T_N$ determined from transport measurements in Fig. 1. Figure 3 shows the rocking scans through the $\textbf{Q}_{AF}$ AF Bragg peak at 28 K, indicating that uniaxial pressure does not change magnetic correlation length.
These results confirm that the pressure evolution of resistivity data is due to bulk properties changes in system and consistent with a non-monotonic suppression of the magnetic order before it vanishes at 280 MPa. %However, given the statistics of the data (Figs. 2 and 3), the results may also be consistent a gradual suppression of the magnetic
%ordered moment with increasing pressure.

To summarize the transport and neutron scattering results, we plot
in Fig. 4(b)
the magnetic ($T_N$) and superconducting ($T_c$) transition temperatures as a function of $P_c$. Both $T_N$ and $T_c$ in Fig. 4(b) are determined from the $d R/d T$ curves of transport data shown in Fig. 1. The $T_N$s determined from our neutron diffraction experiments in Fig. 2(a, b) have also been plotted, and their differences to the transport
measurements can be attributed to the
accuracy of neutron measurements and/or tiny differences in phosphorous concentrations
between the two crystals used in these measurements.
With increasing pressure, $T_N$ approximately
decreases continuously for $P_c>20$ MPa before vanishing abruptly at $P_c = $ 280 MPa
where optimal superconductivity with $T_c \approx 30$ K is achieved.
%Upon further increasing $P_c$, the AF order reemerges
%at the expense of superconductivity, reminiscent of the pressure-induced AF order in optimal doped BaFe$_2$(As$_{0.70}$P$_{0.30}$)$_2$ \cite{Ding2018}.
Simple linear fits to the data for $20$ MPa $<P_c< 280$ MPa yield a reduction in $T_N$ of -25 $\pm$ 3 K/GPa.
For comparison, we note that for optimal $x = 0.30$ compound, a $c$-axis pressure can induce
$T_N$ increase at the rate of 48 $\pm$ 2 K/GPa \cite{Ding2018}, suggesting the sensitivity of magnetism
to quantum fluctuations near optimal superconductivity in BaFe$_2$(As$_{1-x}$P$_x$)$_2$.
Although neutron diffraction measurements
in the high-pressure regime of $P_c>100$ MPa is desirable to confirm the
suppression of $T_N$ to zero attributed from the disappearance of the dip in $d R/d T$ in transport measurements, such experiments are rather difficult for the in-situ uniaxial pressure device used in the neutron diffraction experiments.
%Nevertheless,
%from the pressured-induced
%$d \rho/d T$ behavior in BaFe$_2$(As$_{0.70}$P$_{0.30}$)$_2$ \cite{Ding2018},
%we attribute the appearance of the dip in $d \rho/d T$ at high pressures to the reemergence of the AF order.

In general, electronic phases in iron pnictides
such as BaFe$_2$(As$_{1-x}$P$_x$)$_2$
are believed to be related with crystal structural parameters including pnictogen height (the height of As/P to the Fe layer), $a$, $c$, and the $c/a$ ratio \cite{bohmer17,David2010,Lee2012,Yin2011}.
Specifically, increasing P-doping level in
 BaFe$_2$(As$_{1-x}$P$_{x}$)$_2$ is linearly associated with decreasing pnictogen height, $a$ and $c$ axis, while the $c/a$ is held constant \cite{kasa10}. For comparison, a $c$-axis pressure,
while decreases the $c$-axis and expands $a$-axis lattice constants,
barely changes the iron-pnictogen height \cite{Ding2018}.
From the $P_c$ dependence of the electronic phase diagram
in Fig. 4(b), we see the non-monotonic evolution of the AF order in BaFe$_2$(As$_{0.72}$P$_{0.28}$)$_2$
before its disappearance around $P_c \approx$ 280 MPa with the appearance of optimal superconductivity.
Although a microscopic origin of such a phase diagram is still unclear,
our results suggest that both the out-of-plane and in-plane magnetoelastic coupling are
important for optimal superconductivity and the appearance of a quantum critical point.

In summary, we have systematically studied the $c$-axis uniaxial pressure evolution of the AF phase and superconductivity in underdoped  BaFe$_2$(As$_{0.72}$P$_{0.28}$)$_2$ superconductor.
With increasing $P_c$, the AF order can be gradually suppressed to zero
around at $P_c = 280$ MPa with the appearance of optimal superconductivity,
after the initial enhancement at $\sim$ 20 MPa.
These results indicate that in addition to isoelectronic
doping and hydrostatic pressure, uniaxial pressure along the $c$-axis can be used as a tuning parameter to manipulate the electronic phases and study in the interplay of magnetism and superconductivity in iron pnictides.

%Previous neutron diffraction found that $c$-axis pressure barely changes the pnictide height, but suppresses the $c$-axis and expandes the $a$-axis. It means the

%In previous experiment on optimal doped BaFe$_2$(As$_{0.70}$P$_{0.30}$)$_2$, the induced stripe AF order has been attributed to a strong in-plane magnetoelastic coupling and the increased nearest-neighbor Fe-Fe distance due to c-axis pressure.

%It is the same situation with optimal doped BaFe$_2$(As$_{0.70}$P$_{0.30}$)$_2$ under $c-$axis pressure, pointing to an universal phase diagram BaFe$_2$(As$_{1-x}$P$_{x}$)$_2$ of  under c-axis pressure, which could suggest that c-axis pressure is a good mean to manipulate the phases in iron pnictides.

%the superconducting transition temperature reaches max   , in a fashion similar with the suppression of magnetic order from phosphorous substitution near optimal doped region.
%Interestingly, the magnetic order revives under higher pressure.

%the $T_c$ reaches maximum after the magnetic transition temperature suppressed to zero at 280 MPa. It's worth noting that, the fashion when magnetic transition vanishes in under c-axis pressure is similar with the way it suppressed by phosphorous, and pointing to the same maximum $T_c$.

%the magnetic transition becomes smooth with 66 MPa and 100 MPa and it's har $T_N$ below

%In order to spell out the evolution superconducting transition (T$_c$), we focus,

%In conclusion, we have revealed the electronic phase diagram dependence of c-axis uniaxial pressure in underdoped  BaFe$_2$(As$_{0.72}$P$_{0.28}$)$_2$.

The neutron scattering work at Rice is supported by the
U.S. NSF Grant No. DMR-1700081 (P.D.). A part of the material synthesis work at Rice is
supported by the Robert A. Welch Foundation Grant No. C-1839 (P.D.). The work at BNU is supported by the Fundamental Research Funds for the
Central Universities.

\end{document}